\documentclass[superscriptaddress,twocolumn]{revtex4-1}

\usepackage[utf8]{inputenc}
\usepackage[T1]{fontenc}

\usepackage{graphicx}
\usepackage[position=top]{subfig}

\expandafter\let\csname equation*\endcsname=\relax 
\expandafter\let\csname endequation*\endcsname=\relax 
\usepackage{amsmath}

\usepackage{amssymb}

\usepackage{mathtools}
\usepackage{adjustbox}

\usepackage{braket}

\usepackage{color, colortbl}
\usepackage{xcolor}
\usepackage{color, colortbl}
\definecolor{Gray}{gray}{0.9}
\definecolor{MathGreen}{HTML}{17BB21}
\definecolor{limegreen}{HTML}{32CD32}
\definecolor{goldenrod}{HTML}{DAA520}
\definecolor{crimson}{HTML}{DC143C}
\definecolor{purple}{HTML}{800080}
\definecolor{greenyellow}{HTML}{ADFF2F}

\usepackage[colorlinks=false, citecolor = blue]{hyperref}

\newcommand{\evo}{\text{evo}}

\usepackage{multirow}

\usepackage{comment}

\begin{document}

\title{Approximating quantum thermodynamic properties using DFT}
\author{K. Zawadzki}
\affiliation{Instituto de Física Teórica, UNESP, ICTP South American Institute for Fundamental Research, Rua Dr Bento Teobaldo Ferraz 271, 01140-070, São Paulo, Brazil }
\affiliation{Department of Physics, Royal Holloway University of London, Egham TW10 0EX, UK}

\author{A. H. Skelt}
\affiliation{Department of Physics, University of York, York YO10 5DD, UK}
\author{I. D'Amico}
\affiliation{Department of Physics, University of York, York YO10 5DD, UK}


\date{\today}

\begin{abstract}
The fabrication, utilisation, and efficiency of quantum technologies rely on a good understanding of quantum thermodynamic properties.  Many-body systems are often used as hardware for these quantum devices, but interactions between particles make the complexity of related calculations grow exponentially with the system size. Here we explore and systematically compare `simple' and `hybrid' approximations to the average work and entropy variation built on static density functional theory concepts. These approximations are computationally cheap and could be applied to large systems.
We exemplify them considering driven one-dimensional Hubbard chains and show that, for `simple' approximations and low to medium temperatures, it pays to consider a good Kohn-Sham Hamiltonian to approximate the driving Hamiltonian. Our results confirm that a `hybrid' approach, requiring a very good approximation of the initial and, for the entropy, final states of the system, provides great improvements. This approach should be particularly efficient when many-body effects are not increased by the driving Hamiltonian.
\end{abstract}


\maketitle
\section{Introduction}

Many-body systems composed by a relatively small number of relevant degrees of freedom are often proposed and used as hardware for quantum devices. This is one of the reasons behind the development of quantum thermodynamics (QTD)\cite{Vinjanampathy2016,Goold2016,Millen2016,Parrondo2015}, because it prevents relying on assumptions derived from the thermodynamic limit.
Understanding thermodynamic properties of these systems is crucial as they could limit applications, but also help the fabrication and running of efficient quantum devices.  In this sense, the average quantum work extracted and the related thermodynamic entropy are  very relevant:  for example, quantum work is important for quantum engines, quantum batteries, and optimal energy consumption, whilst the thermodynamic entropy (or irreversible work) is indicative of the energy dissipated in a cycle or to reset a system to thermal equilibrium.

The dynamics of interest to perform thermodynamic cycles or computational algorithms is often a rich and non-trivial out-of-equilibrium dynamics. 
Recent studies looked at work and entropy in many-body quantum harmonic oscillators and spin chains \cite{Silva2008,Dorner2012,Joshi2013,Mascarenhas2014,Sindona2014,Fusco2014,Zhong2015,Eisert2015,Bayat2016,Solano-Carrillo2016}. 
However, various systems of interest are non-integrable and calculating the systems' properties may become challenging even for integrable systems, for instance, in scenarios where an external drive breaks integrability. Even well established methods for strongly correlated systems quickly reach a bottleneck. This is more critical at finite temperature, when properties of interest involve excited states. 
For instance, the Density Matrix Renormalization-Group, which yields almost exact results for ground-state and time-dependent  properties \cite{FEIGUIN-real-t-DMRG}, has limited applicability at finite temperature \cite{FEIGUIN-finite-T-DMRG}. Accuracy is lost whenever the evolution in imaginary time has to run for longer times than the inverse of the correlation energy scales.  Therefore, finding a way to accurately incorporate interactions into reliable approximations is an important issue for quantum thermodynamics.

Recently, we proposed a scalable approximation protocol inspired by density functional theory (DFT) \cite{Herrera2017,Herrera2018} and tested it to calculate the quantum work in a Hubbard dimer driven at finite time.  
Later, in reference  ~\cite{Skelt2019JPA} we quantified the error in neglecting the Coulomb coupling in Hubbard chains. We demonstrated that increased accuracy for estimating quantum work can be achieved in a wide region of the parameter space by means of a hybrid approximation which uses the exact (or highly accurate) initial state, whilst constructing the evolution operator from the non-interacting Hamiltonian.  
This approach is numerically relatively cheap. However, it leads to the question if a more sophisticated hybrid approximation, including many-body interactions when calculating the system evolution, for instance via Density Functional Theory (DFT), would yield results of greater accuracy. 

DFT\cite{Gross2013density, Capelle2006bird} is one of the most successful methods to derive properties of complex many-body systems and materials, providing a suite of approximations relevant to continuous electronic systems and model Hamiltonians \cite{Kurt-Scho.PhysRevB.52.2504,Capelle2013}.

The development and applications of DFT to finite temperature problems \cite{Querne2019finite}, and especially to finite-temperature systems out-of equilibrium, are comparatively still in their infancy. Hence the importance of the present study, in which we combine and adapt some well-known zero-temperature, DFT approximations for model Hamiltonians with the `hybrid' approach from reference~\cite{Skelt2019JPA} and test their limits in describing the quantum thermodynamics of finite-temperature systems strongly out of equilibrium. This work will complement and extend the recent work~\cite{Rehra2018} which considered static systems and showed that approximations to the exchange-correlation potential built for ground-states should be accurate at finite $T$ below a characteristic thermal scale associated with the Fermi energy.
Importantly, the DFT-based set of approximations that we propose, will remain applicable to systems of higher complexity, as they remain numerically relatively inexpensive.

\section{Method}

\subsection{Average quantum work and entropy production}
The average quantum work quantifies the energy lost or acquired by a quantum system in a controllable way during a certain dynamics. For closed systems where the initial state $\rho_0$ is diagonal in the energy basis, it is equal to \cite{Vinjanampathy2016}
\begin{equation}
\label{eq:q_work_trace}
\langle W \rangle = \mathrm{Tr}\left[ \rho_{f} \hat{H}_{f} \right] - \mathrm{Tr}\left[ \rho_0 \hat{H}_0 \right],
\end{equation}
where $\rho_f$ is the system state at the final time. The work that can be extracted from a system is then $\langle W_{ext} \rangle= - \langle W \rangle$.

The entropy production is related to the energy that would have to be dissipated for the system to return to equilibrium after a thermodynamic process \cite{Herrera2017,Batalhao2015}.  
In closed quantum systems, the {\color{black} non-equilibrium entropy} can be defined as \cite{Deffner-PhysRevLett.105.170402,Batalhao2015,Goold2016}:
\begin{eqnarray}
\label{eq:entropy}
\Delta S &=&  \beta \left( \langle W \rangle - \Delta F \right) \\
&=&  - \beta \left( \langle W_{ext} \rangle + \Delta F \right)
\label{eq:entropy_ext}
\end{eqnarray}
where $\beta = 1/k_B T$, and the free energy variation is
\begin{equation}
\label{eq:free_energy}
\Delta F = - \frac{1}{\beta} \ln \left( \frac{Z_f}{Z_0} \right),
\end{equation}
with $Z_0$ the partition function for the initial, and $Z_f$ for the final Hamiltonian.

\subsection{Model system}
\label{sec:params}

We will apply the approximations proposed to the
one-dimensional Hubbard model\cite{Essler2005, LIEB-PA.321.1, GOHMANN-PLA.263.4}. It is used to simulate structures such as chains of atoms or coupled quantum dots \cite{Coe2010,Yang2011,Murmann2015,Coe2011,Brown2019,Nichols2019}, which are relevant as hardware for quantum technologies. It also allows representation of a wide range of phases of matter: metallic, Mott-insulating, band-insulating \cite{Manmana_2004}, and even superconducting \cite{Essler2005, Lecheminant-PhysRevLett.95.240402, Feiguin-PhysRevB.79.100507, Franca2006,Kohno-PhysRevLett.105.106402, Franca2012,dePicoli2018}.

Calculating the thermodynamical properties of Hubbard chains in the presence of inhomogeneities is not possible analytically;  numerically, achieving reasonable accuracy requires  substantial computational effort. The size of their Hilbert space increases exponentially with the number of sites. At finite temperature and/or out-of-equilibrium dynamics, more excited states are populated.  Hence the calculation 
of work and entropy, becomes prohibitive for systems exceeding a dozen sites. For larger systems, the simulation of long time evolutions suffers from propagation of errors, and becomes especially problematic at time scales of the order of $10^2$ the inverse of the hopping parameter \cite{KENNES201637, Goto2019, krumnow2019towards}. 

Yet, in short Hubbard chains non-trivial behaviors are already evident, such as the precursor to the metal-Mott insulator phase transition and the transition to a band-insulator phase. 
 This makes these systems ideal to test approximations for quantum thermodynamics\cite{Herrera2018,Herrera2017,Carrascal2015,Fuks2014}.
 
The Hamiltonian for a chain of $N$ sites is
\begin{multline}
\label{eq:Hubbard_Hamiltonian}
{\hat{H}(t)} = -J \sum_{i,\sigma}^N \left( \hat{c}^{\dagger}_{i,\sigma} \hat{c}_{i+1,\sigma} + \hat{c}^{\dagger}_{i+1,\sigma} \hat{c}_{i,\sigma} \right) \\ + U \sum_i^N \hat{n}_{i,\uparrow} \hat{n}_{i,\downarrow} + \sum_i^N v_i(t) \hat{n}_i,
\end{multline}
where $J$ is the hopping term, $U$ the Coulomb interaction strength on site $i$, and $v_i(t)$ is the time-dependent driving potential at site $i$.  The creation (annihilation) operator  for a fermion with spin $\sigma$ ($\sigma= \uparrow$ or $\downarrow$) at site $i$ is $\hat{c}^{\dagger}_{i,\sigma}$ ($\hat{c}_{i,\sigma}$), and its number operator is $\hat{n}_i = \hat{n}_{i,\uparrow} + \hat{n}_{i,\downarrow}$, where $\hat{n}_{i,\sigma} = \hat{c}^{\dagger}_{i,\sigma} \hat{c}_{i,\sigma}$. The number of particles is $N$, with $n_\uparrow=n_\downarrow$, and we consider Open Boundary Conditions (OBC).
{\color{black} OBC provide further inhomogeneity to the chain and can be experimentally implemented \cite{PhysRevLett.121.130402,Trotzky2012probing,kohlert2021experimental,Hensgens2017quantum}}.

The external potential $v_i(t)= \mu_i^0 + \mu_i^\tau t / \tau$  makes the system inhomogeneous \footnote{We note that also OBC makes the system inhomogeneous. At half-filling, the density is homogeneous if the number of sites is even and if there are no local inhomogeneous potentials.\cite{Zawadzki2017twist}}.  Here $\mu_i^0$ and $\mu_i^{\tau}$ are the time-independent coefficients for site $i$ at the initial time $t=0$ and final time $t=\tau$, respectively.
We note that the final Hamiltonian is independent from $\tau$, which then sets the rate of driving during the system evolution. 

We consider systems with $N=6$; for each site $i$, $\mu_i^0 = \mu_0 (-1)^i$ at $t=0$ and $\mu_i^{\tau} = \mu_{\tau} (-1)^i$ at $t=\tau$, where $\mu_0 = 0.5J$ and $\mu_{\tau} = 4.5J$ (`Comb' potential~\cite{Skelt2019JPA}). 
These chains and dynamics are directly comparable to the ones in reference \cite{Skelt2019JPA}

\subsubsection{Metallic, Mott insulator and band insulator phases}
equation~\ref{eq:Hubbard_Hamiltonian} with the `comb' potential describes an ionic Hubbard model (see e.g. \cite{Manmana_2004}), with the ratio between $U$ and the time-dependent height of the local alternating potential  changing with time.
For an adiabatic dynamics, for $U\lesssim J$, as $t$ increases the system changes from metallic to a quasi-band-insulator, while for $U\gg J$ and $U\lesssim 2|v_i(\tau)|$ it would transition from a quasi-Mott-insulator to a quasi-band-insulator.
For a generic dynamics, if the system will or not settle in each of these  phases will depend on the rate of the finite-time dynamics.

The presence of these phases represents a tough test for the approximations analyzed in this paper.

\subsection{Density functional theory approach to quantum thermodynamics}
\label{subsec:DFT_QTD}

DFT is, in principle, an exact method of calculating
properties of many-body systems. In fact, in 1964 Hohenberg and Kohn
, demonstrated that any property of a ground state system can be, in theory, expressed as a functional of the ground state density\cite{Hohenberg1964}, and hence calculated if the functional is known or can be reasonably approximated. Afterwards, the method was extended to spin-dependent~\cite{Fran_a_2012,Vieira2014spinchargesep} and time-dependent problems~\cite{Ullrich2013}, which also allows  the calculation of excited states. Extensions to mixed states and finite temperature systems are also pursued, with some success~\cite{Vivaldo-PRA.92.013614, Querne2019finite, Burke-PhysRevB.93.195132}.
Often the wavefunction of complex many-body systems is too large to handle, and/or the related (time-dependent) Schr\"odinger equation too complex to solve; through DFT, their properties can be estimated using their local densities, which are much simpler quantities both to compute and to measure.

To find the ground state density of a given system, Kohn and Sham developed a method which mapped the interacting system to a fictitious non-interacting system whose potential - the Kohn-Sham (KS) potential - is constructed in such a way to reproduce the many-body interacting ground state density \cite{Kohn1965}.  In practice, finding this density  through this KS system usually requires some approximation to be made in the so-called exchange-correlation potential, the key ingredient of the KS potential.  This has led to many DFT approximations~\cite{Capelle2006,Capelle2013,Su2017,Mardirossian2017}.

Among them, there are  approximations suitable to describe Hubbard chains in the static zero-temperature case, see e.g. \cite{Capelle2013}.  In the next two subsections, we review two of them and explain how we adapt them to finite temperature; we then explain how  thermodynamic quantities can be calculated based on them.

\subsubsection{BALDA} One of the most commonly used DFT approximation is the local density approximation (LDA). This is designed to represent spatially slowly varying continuous systems \cite{Kohn1965,Entwistle2016}, but it is often successful well beyond this limit. Notably, LDA is computationally `cheap' to implement even for very large systems.  An extension of LDA specifically designed for the Hubbard model is the Bethe Ansatz LDA (BALDA) developed by Lima \textit{et al.}\cite{Lima2003}.
The BALDA KS potential $v^{BALDA}_{KS,i}$ is
\begin{equation}
    v^{BALDA}_{KS,i} = v_{i} + v^{BALDA}_{H,i} + v^{BALDA}_{xc,i}.
\label{vKS_BALDA}
\end{equation}
Here the external potential $v_{i}$ is the same as applied to the original many-body system. To adapt the scheme to finite temperatures, we have constructed the Hartree
term  as $v^{BALDA}_{H,i}=Un_{0,i}^{th,exact}/2$, where $n_{0,i}^{th,exact}$ is the occupation at site $i$ obtained from the exact initial thermal state. Our exchange-correlation terms  $v^{BALDA}_{xc,i}$ also utilizes $n_{0,i}^{th,exact}$, and it is formally defined in~\cite{Lima2003}.

In \ref{app_BALDA_temp} we show the effect of using a thermal density in this approximation for the initial local densities, the initial KS potential, and the instantaneous Hamiltonian spectrum. We note that the effect on the spectrum is small.


\subsubsection{Exact ground state Kohn-Sham potential}

Refs~\cite{Smith2016,Wagner2014,Rehra2018} have started to explore reverse engineering the thermal density to find the Kohn-Sham system at  temperatures which are much less than the electron-electron interaction strength. It has been shown that the exact Kohn-Sham system at low temperatures is remarkably similar to the ground state (zero temperature) Kohn-Sham system.  Here, we will test if the {\it dynamics} at low temperatures can be accurately reproduced by an evolution operator built upon the exact ground state (GS) Kohn-Sham potential $v^{GS}_{KS,i}$, $i=1,N$. We will also check its limits as the temperature increases.

We employ the reverse engineering scheme developed in reference~\cite{Coe2015}, which is suitable for ground state lattice systems at zero temperature. In the present work, the initial-time GS exact density  $n_{0,i}^{GS,exact}$ is used as the input of a self-consistent cycle that yields the exact Kohn-Sham potential at $t=0$ as the output.  

The Kohn-Sham potential $v^{GS}_{KS,i}$ is
\begin{equation}
    v^{GS}_{KS,i} = v_{i} + v^{GS}_{H,i} + v^{GS}_{xc,i},
\label{vKS_GS}
\end{equation}
where the external potential $v_{i}$ is the same as applied to the original many-body system; the Hartree term is $v^{GS}_{H,i}=Un_{0,i}^{GS,exact}/2$,   and the exchange-correlation terms $v^{GS}_{xc,i}$ is defined by the reverse engineering result.

We note that with the definitions chosen, $v^{GS}_{H,i}\ne v^{BALDA}_{H,i}$ for $T\ne 0$, and the comparison between the `BALDA' and `GSKS' approximations becomes a comparison over contributions from both the Hartree and the exchange-correlation potentials.

\subsubsection{Simple and Hybrid DFT approximations for quantum thermodynamics}
\label{sec:defining_DFT_approxes_QT}

\begin{table*}[ht!]
{\color{black}
    \caption{
    {\color{black}`Simple' (first and third) and `hybrid' (second and fourth) proposed approximations for calculating the systems' dynamics and hence to estimate the average work from equation~\ref{eq:q_work_trace}.
    Here   $\hat{H}^{\evo}(t)$ is the Hamiltonian which defines the time-evolution operator; initial states are indicated. $Z^{Acronym}$ is the partition function calculated using $\hat{H}^{Acronym}$, with $Acronym$ = $BALDA,~GSKS,~exact$.
    }
    }
    \label{tab:approxes_DFT}
    \centering
    \begin{tabular}{|c|c|c|}
        \hline 
         Approximation & Initial State &  $\hat{H}^{\evo}(t)$ \\
         \hline
         $\langle W ^ {BALDA} \rangle $ & 
         $\hat{\rho}^{BALDA}_0 = \frac{\exp \left( -\beta \hat{H}^{BALDA}(0) \right)}{Z^{BALDA}}$ & 
         \multirow{3}{*}{
         $\displaystyle \hat{H}^{BALDA} =  -J \sum_{i,\sigma}^N \left( \hat{c}^{\dagger}_{i,\sigma} \hat{c}_{i+1,\sigma} + h.c. \right) + \sum_i^N  v^{BALDA}_{KS,i}(t) \hat{n}_i $ 
          }\\
         & & \\
         $\langle W ^ {exact+BALDA} \rangle $  & $\hat{\rho}^{exact}_0 = \frac{\exp \left( -\beta \hat{H}(0) \right)}{Z^{exact}}$ &  \\
         \hline
         \hline
         $\langle W ^ {GSKS} \rangle $ & 
         $\hat{\rho}^{GSKS}_0 = \frac{\exp \left( -\beta \hat{H}^{GSKS}(0) \right)}{Z^{GSKS}}$ & 
         \multirow{3}{*}{
         $\displaystyle \hat{H}^{GSKS} =  -J \sum_{i,\sigma}^N \left( \hat{c}^{\dagger}_{i,\sigma} \hat{c}_{i+1,\sigma} + h.c. \right) + \sum_i^N  v^{GS}_{KS,i}(t) \hat{n}_i $ 
          }\\
         & & \\
         $\langle W ^ {exact+GSKS} \rangle $  & $\hat{\rho}^{exact}_0 = \frac{\exp \left( -\beta \hat{H}(0) \right)}{Z^{exact}}$ &  \\
         \hline
    \end{tabular}
}
\end{table*}

{\color{black}
Reference~\cite{Skelt2019JPA} established that using the exact initial state with dynamics approximated using the non-interacting Hamiltonian yields results of surprisingly high accuracy, even up to strong many-body interaction strengths. There, `simple' and `hybrid' approximations for the average work and entropy variation are recast in the following forms

\begin{equation}
\label{eq:q_work_trace_approx}
\langle W^{is+\evo} \rangle = \mathrm{Tr}\left[ \rho_{f}^{is+\evo} \hat{H}^{\evo}(t=\tau) \right] - \mathrm{Tr}\left[ \rho_0^{is} \hat{H}^{\evo}(t=0) \right],
\end{equation}

and

\begin{equation}
\label{eq:entropy_approx}
\Delta S = \beta (\braket{W^{is+\evo}} - \Delta F^{is} ).
\end{equation}

where $is$ (initial system) refers to the approximation used to derive the initial state,  $\rho_0^{is} = \exp \left( -\beta \hat{H}^{is}(0) \right)/ \mathrm{Tr}\left[ \exp \left( -\beta \hat{H}^{is}(0) \right) \right]$, and $\evo$ is the approximation used for the evolution operator $\mathcal{U}_{\evo}= \mathcal{T} e^{-i \int_0^{\tau}\hat{H}^{\evo}(t) dt}$ where $\mathcal{T}$ is the time-ordered operator.   The final state is then  $\rho_{f}^{is+\evo} = \mathcal{U}_{\evo} \rho_0^{is} \mathcal{U}_{\evo}^{\dagger}$. In the `simple' approximations, $is=evo$ and only $is$ is written. 
Where Ref.~\cite{Skelt2019JPA} utilized  a non-interacting approach, here we will use either the BALDA or the GSKS approximations. 
The related implementations are summarised in table~\ref{tab:approxes_DFT}.
}

In all cases, the (approximate) Hartree and the (approximate) exchange correlation potential are {\it time-independent}, which keeps the computational cost low. Within this framework, the BALDA-based approximations represent a lower bound to the accuracy that can be achieved, whilst the approximations based on the $t=0$ reverse-engineered Kohn-Sham potential  give an estimate of its upper bound.

As shown in table~\ref{tab:approxes_DFT}, for both the `simple' (second column) and the 'hybrid' (third column) approximations, we include interactions in the driving Hamiltonian $\hat{H}^{\evo}(t)$ through the approximations of the Kohn-Sham potential as described in the previous subsections. The KS potentials are made time dependent by the substitution $v_{i}\to v_{i}(t)$ in equations (\ref{vKS_BALDA}) and (\ref{vKS_GS}), with $v_{i}(t)$ the `comb' potential previously described.
In the `hybrid' approximations, interactions are also included by utilizing the exact
initial thermal state, under the assumption that this is
relatively easy to calculate (or accurately estimate)  for a static system at equilibrium.

As the exchange-correlation and Hartree potentials are only calculated at $t=0$ as derived from the ground state theory, calculations remain ``cheap''. This is still a relatively crude approximation to the many-body dynamics, but constitutes the next step towards including interactions within the system dynamics with respect to Reference~\cite{Skelt2019JPA}.

\section{Results}

We will now present the results for the average quantum work from the four approximations in table~\ref{tab:approxes_DFT},  including their comparison with the corresponding exact results and non-interacting approximations reported in ~\cite{Skelt2019JPA}. The exact results are obtained by exact diagonalization  and propagation~\cite{Qutip-paper} of  the system Hamiltonian in equation (\ref{eq:Hubbard_Hamiltonian}).

Three temperatures are considered: a low temperature  ($T=0.2J/k_B$), a medium temperature  ($T=2.5J/k_B$), and a high temperature ($T=20J/k_B$).  For each of them, we explore regimes from non-interacting ($U=0J$) to strongly interacting ($U=10J$), and from sudden quenches ($\tau=0.5/J$) to quasi-adiabatic evolutions ($\tau = 10/J$).

\subsection{Simple BALDA approximation}

\begin{figure*}
\centering
\includegraphics[width=0.95\textwidth]{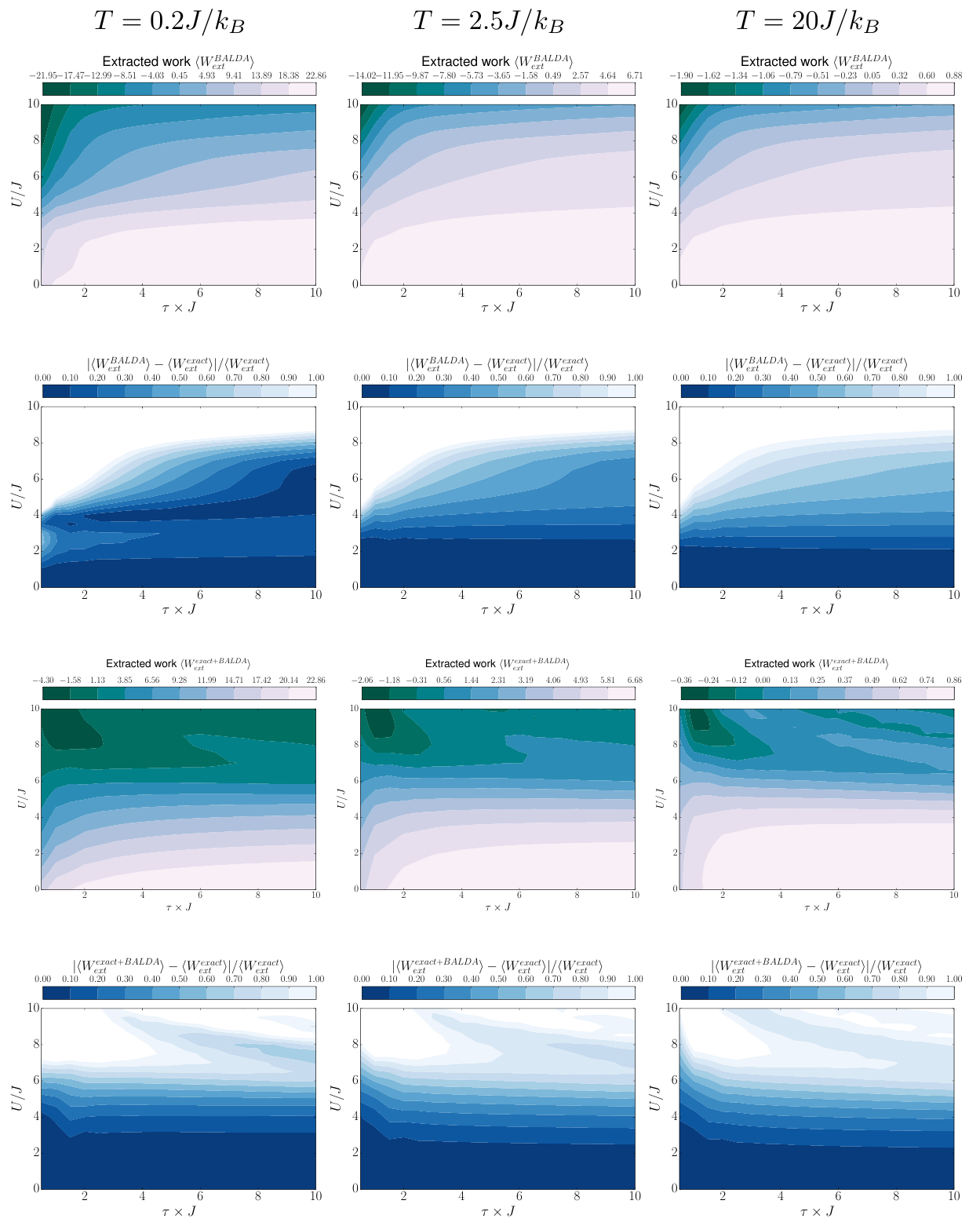}
\caption{First row: Work extracted using simple BALDA for $0.5 \leq \tau \times J \leq 10$ ($x$-axis) and $0 \leq U/J \leq 10$ ($y$-axis) for 6 site chains with `comb' potential. Temperature increases from the left to the right column; lighter shades correspond to greater work extracted.
Second row: Relative difference between simple BALDA and exact results for the work.  Same parameters as first row; darker shades correspond to higher accuracy.
Third and fourth rows: same as in first and second row, but for hybrid BALDA.
}
\label{fig:BALDA_work}
\label{fig:ex+BALDA_work}
\end{figure*}

Figure~\ref{fig:BALDA_work} shows the results of the simple BALDA approximation for the average quantum work (top row) and its relative error (second row) to the exact calculation. The temperature grows from left to right. In the top panels, the work that can be extracted ($W>0$) increases with lighter shades. In the bottom panels,the relative error increases with lighter shades.

\begin{figure*}[htb!]
\centering
\includegraphics[width=0.85\textwidth]{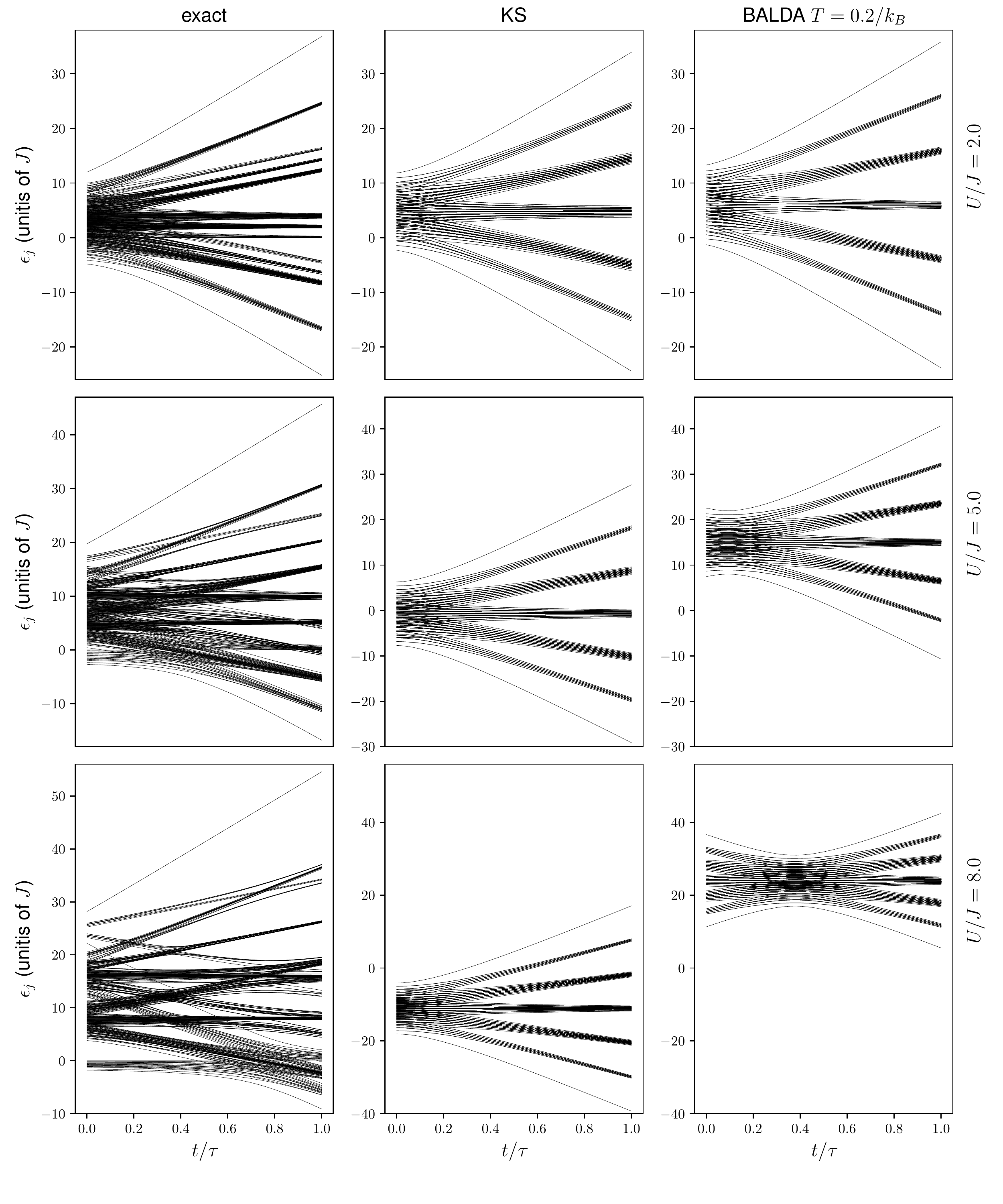}

\caption{Upper panels: exact  (left), simple and hybrid GSKS (centre), and  simple and hybrid BALDA (right) instantaneous spectra for $U=2J$. Central and lower panels: as for upper panels, but for $U=5J$ and for $U=8J$, respectively. Here the low-temperature simple-BALDA is used, but differences with higher temperatures are small, see \ref{app_BALDA_temp}.}
\label{fig:simple_spectrum}
\end{figure*}

\begin{figure*}[htb!]
\centering
\includegraphics[width=0.9\textwidth]{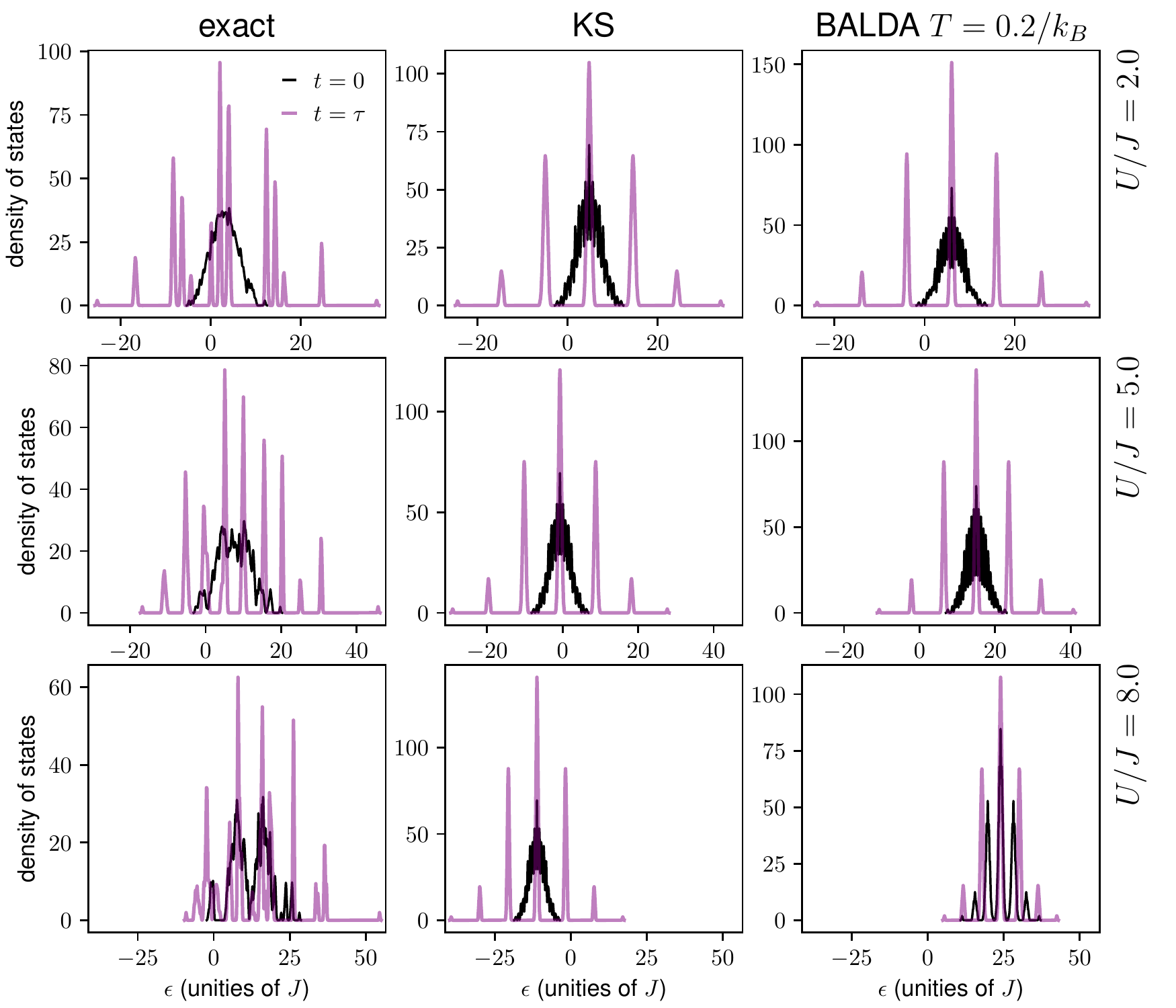}

\caption{Density of states (DoS) for $U=2J$ (top) $U=5J$ (middle) and for $U=8J$ (bottom) at $t=0$ (black) and $t=\tau$ (purple); first column: exact; second column: simple GSKS; third column: simple BALDA. Here the low-temperature simple-BALDA DoS is plotted.}
\label{fig:DoS}
\end{figure*}

As mentioned, BALDA assumes a slowly varying infinite system. Here we have short chains but with a slowly varying potential at the beginning of the dynamics, when the BALDA KS potential is calculated. This condition is not satisfied as the dynamics progresses due to the external drive.

As $U$ increases, double site occupation is discouraged, and a charge gap opens up in the lower part of the spectrum, bringing the system towards a Mott insulating state (see figure~\ref{fig:simple_spectrum}, left column, for $t/\tau \approx 0$).  This feature is embedded in the static BALDA exchange correlation potential, which displays a discontinuity at half filling that becomes more pronounced as $U$ increases. This positively reflects in the simple-BALDA instantaneous spectra (figure~\ref{fig:simple_spectrum}, right column, $t/\tau \approx0$), where we can see that, as $U/J$ increases, the 20 eigenstates without contributions from double site occupation gets separated by a charge gap from the rest of the spectrum.
However, as $U/J$ increases, the {\it quantitative} energy-level spacing becomes quite different, and hence it does the level population at the beginning of the dynamics. As a result, the `simple BALDA' approximation is only able to quantitatively capture the work accurately for Coulomb repulsion of the order of $U\stackrel{<}{\sim} 3 J$, performing well for a metallic system with an accuracy of $\sim 20\%$ or better almost at all $\tau$'s.

When $U$ and $v(t)$ become comparable, a precursor to a transition to a band-insulator starts to appear in the instantaneous spectrum (see first column of figure~\ref{fig:simple_spectrum} as $t/\tau$ increases).  This is in general well reproduced by the lower part of the simple-BALDA spectrum (see third column of figure~\ref{fig:simple_spectrum}). When $U\stackrel{>}{\sim}5J$, as $t/\tau$ increases, the exact instantaneous spectrum displays a crossover between the Mott-insulating and the doublon-hole correlation phase \cite{Hartke-PRL.125.113601}  driven by the applied field. As mentioned, simple BALDA does not reproduce quantitatively well the phase dominated by many-body interactions, so that for $U\stackrel{>}{\sim}5J$ accuracy reduces greatly for most $\tau$'s at all temperatures.

Finally, for simple BALDA and $U \gtrsim 6J$, work needs to be performed \textit{on} the system - note the change in sign - to achieve the driven dynamics for all temperatures: this is an artifact of simple BALDA, as the extracted work is always positive for the exact results~\cite{Skelt2019JPA}. 

For $U\gtrsim 5J$ the width of the BALDA spectra is significantly smaller than the corresponding exact ones, see figure~\ref{fig:simple_spectrum}, contributing to significant differences in the corresponding density of states, see figure~\ref{fig:DoS}.  As the temperature increases, the above leads to the incorrect description of the populations of the approximated initial thermal state, and poor performance of simple BALDA for medium-large $U/J$ and temperatures. 

However, we note that the maximum work extracted, which is achieved for very weakly correlated regimes, is accurately estimated by BALDA, especially at low and intermediate temperatures ~\cite{Skelt2019JPA}.

\subsection{Hybrid BALDA approximation}

Figure~\ref{fig:ex+BALDA_work} shows the results for the average work extracted using the hybrid BALDA approximation (third row), and the relative difference with the exact results (fourth row) for our parameter set.  By including the exact initial state, the quantitative accuracy is similar across all temperatures, with a marked improvement at low temperatures and $U\stackrel{<}{\sim} 3J$, where the accuracy is now consistently $10\%$ or less. Also, for all temperatures and small $\tau$'s,  $\tau \stackrel{<}{\sim} 1$ (quasi-sudden quench regime), the accuracy is consistently improved for $3J\stackrel{<}{\sim} U\stackrel{<}{\sim} 6J$. The extracted minimum work  is also better captured than with simple BALDA, though the minimum work within the considered  parameter space is still negative. We will further comment on the reasons for this improvements in section \ref{simpvshybr}. Comparison with the corresponding exact heat maps in ~\cite{Skelt2019JPA} shows that hybrid BALDA produces qualitatively incorrect trends with $\tau$ and $U$ for  $U\stackrel{>}{\sim} 7$.

\subsection{Simple GSKS approximation}

\begin{figure*}
\centering
\includegraphics[width=\textwidth]{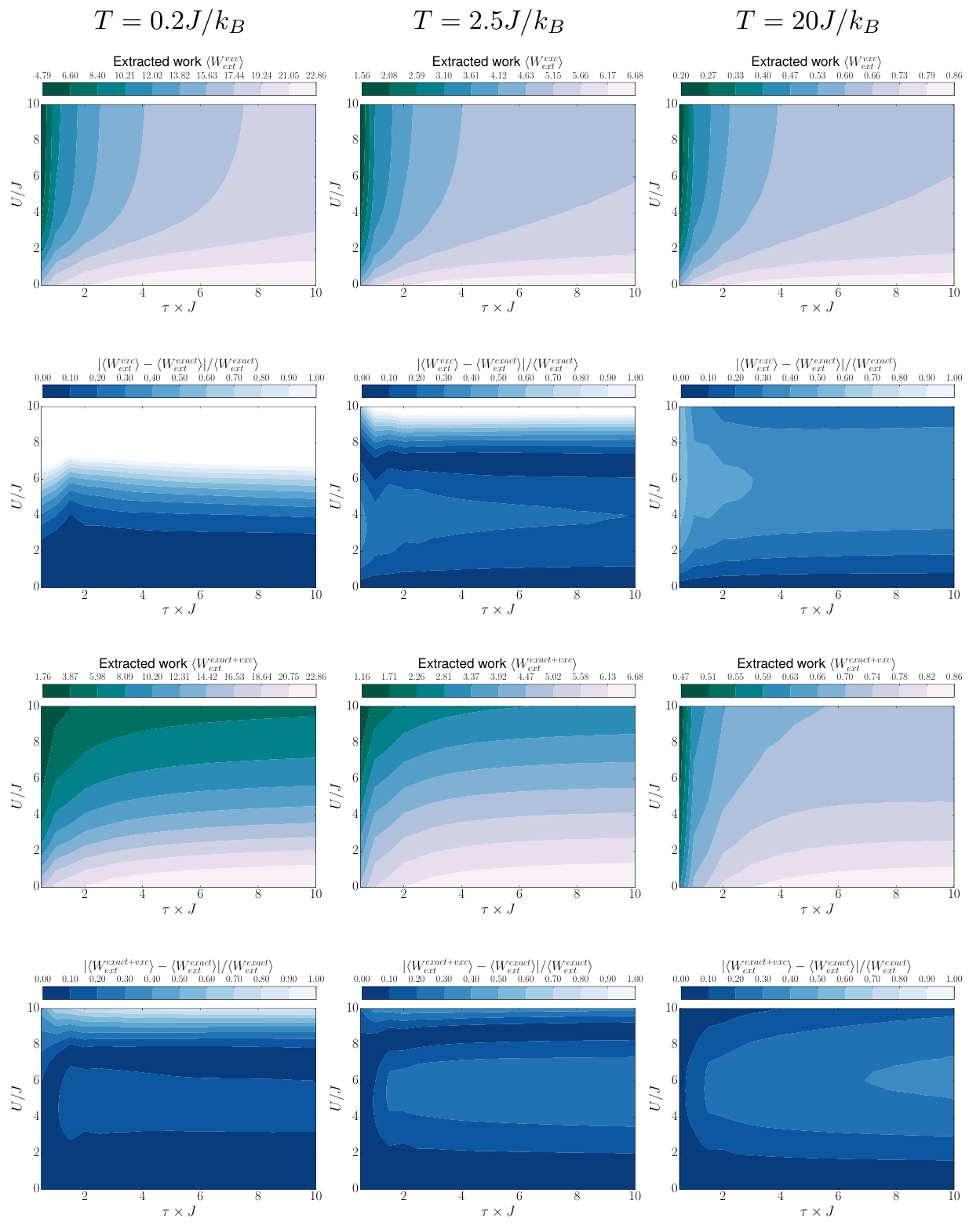}
\caption{Same as in figure~\ref{fig:BALDA_work} but for GSKS (first and second rows) and hybrid GSKS (third and fourth rows).} 
\label{fig:vxc_work}
\label{fig:ex+vxc_work}
\end{figure*}

As mentioned, this approximation is based on the exact ground state $\hat{v}_{xc}$ as calculated at $t=0$ using the inversion scheme of reference~\cite{Coe2015}. The results for the extracted work (top row) and relative difference (second row) with the exact results are shown in Fig \ref{fig:vxc_work}. For $U\stackrel{>}{\sim} 5$, results are almost independent from many-body interactions, while at lower $U$'s results are qualitatively more similar to the exact ones for low and medium temperatures.

At the lowest temperature, the simple GSKS approximation  improves over the simple BALDA results for $1\stackrel{<}{\sim}U\stackrel{<}{\sim} 3$: simple GSKS reproduces a bit better the exact instantaneous spectrum at low $U$ values, such as the low-energy levels and related first gap, see upper panels of figure~\ref{fig:simple_spectrum}. Interestingly, as $U$ increases, the simple GSKS instantaneous spectrum does not reproduce the charge gap-opening due to the precursor to the Mott-insulator transition which is visible in both exact and simple BALDA spectra (see lower panels of figure~\ref{fig:simple_spectrum} for low $t/\tau$ values). This lack of structure will negatively affect results especially at higher temperatures, when states above the charge gap could become substantially occupied. Indeed, at medium and high temperatures the region with accuracy of $\lesssim 20\%$ with respect to the exact shrinks with respect to simple BALDA.
For $T=20J/k_B$, accuracy is reduced to within $\sim 50\%$ of the exact result for the majority of $U$'s and $\tau$'s, which is actually worse than the completely non-interacting approximation `NI' (figure 4 of~\cite{Skelt2019JPA}).  At this temperature interactions are almost negligible for  $U \lesssim 5J$, while this is the region where simple GSKS shows dependence on $U$.

These results expose the limits of the simple GSKS approximation, where $v_{xc}$ is accurately constructed for the specific system albeit at low temperatures, where it performs better than simple BALDA. However we note that this GSKS approximation captures the minimum work extracted over the parameter set much more accurately than either simple or hybrid BALDA approximations. In particular, there is no negative work extraction in the GSKS results for the `comb' potential at 6 sites, much like in the exact results.

\subsection{Hybrid GSKS approximation}

Finally, results for the hybrid-GSKS approximation are shown in Figure~\ref{fig:ex+vxc_work}, third and fourth rows. 
We observe qualitative \textit{and} quantitative improvement in the accuracy over the other three approximations shown in this work.  Qualitatively behaviour is now well reproduce at all $U$'s for low and intermediate temperature and at medium-large $U$'s for large temperature.
Estimates are within 20-30\% of the exact results over most regimes for all temperatures.  Notably, there is a marked improvement in the strongly-correlated quasi-sudden quench regime.
Also, both the minimum and maximum values of average work extraction across the parameter set are captured well, usually within 20\% of the exact value.

\subsection{Comparison with the approximations of reference \cite{Skelt2019JPA}}
\label{subsec:comparison}

As mentioned, reference \cite{Skelt2019JPA} explored `simple' and `hybrid' approximations in which a fully 
non interacting evolution Hamiltonian  $H^{\evo}(t)= H(t;U=0)$ is considered. The hybrid (`exact+NI') approximation was found to perform surprisingly well in estimating results for the average work. Now, we can discuss how the DFT-inspired approximations proposed in the present work compare to the results of this previous work.

For low temperatures, simple BALDA shows an improvement over the completely non-interacting approximation `NI' from reference~\cite{Skelt2019JPA}, figure 4. However, this improvement is lost as the temperature increases.  This could be attributed to two main factors. First, BALDA is an approximation for the ground state (and hence zero temperature): as the temperature increases an increasing number of higher energy states are populated and so a ground state approximation is less reliable. Second, as $T$ increases, thermal excitations allow electrons to move almost freely among levels of energy, so that a non-interacting approximation becomes increasingly more suitable.

Surprisingly the hybrid BALDA approximation, where interactions are somewhat included in $H^{\evo}$, performs worse than the hybrid NI approximation (see figure 5 from reference~\cite{Skelt2019JPA}). 
The NI spectrum occupy an energy window similar to the exact spectrum for $U=2J$. 
Note then that the window of energy in which NI and the $U=8J$ exact eigenstates exist are closer to each other than the $U=8J$ spectra for the exact and BALDA cases.

 In addition, the improvement from BALDA to hybrid BALDA is modest when compared to the improvement found from simple NI to hybrid NI~\cite{Skelt2019JPA}.  This suggests that to improve over the hybrid NI approximation by incorporating electron-electron interactions into the evolution Hamiltonian, one should use either a static DFT approximation more elaborate than BALDA, or perhaps include a time-dependency in the exchange-correlation potential as proposed in \cite{Herrera2017,Herrera2018}.

At all temperatures, for $U\stackrel{>}{\sim} 5$,  the simple GSKS approximation is almost independent from many-body interactions, so its results are qualitatively similar to the corresponding NI results in figure 4 of~\cite{Skelt2019JPA}. However, at the lowest temperature explored, the simple GSKS approximation substantially improves over its NI counterpart for $1\stackrel{<}{\sim}U\stackrel{<}{\sim} 5$. For $T=20J/k_B$, simple-GSKS accuracy for the majority of $U$'s and $\tau$'s is actually worse than the simple NI approximation (figure 4 of~\cite{Skelt2019JPA}).  This confirms the limitations of simple-GSKS at higher temperatures, as this approximation is constructed from ground state properties.

Results from hybrid GSKS and hybrid NI approximations are comparable (see figure 5 of \cite{Skelt2019JPA}), with very similar patterns for the average work approximation and its relative error. For the low and medium temperatures, the hybrid GSKS approximation is better at low $\tau$ ($\tau\lesssim 1$) and increasing coupling strength $U$. For medium and high temperatures, it is generally less accurate than the hybrid NI approximation for $\tau\stackrel{>}{\sim}1$ and medium interaction strengths.  These differences can be attributed to the dominant energy at play in each situation: in very strongly interacting systems at low and medium temperatures, $U$ will be a dominant energy and so the inclusion of a system-specific accurate $v_{xc}$ will help capture these interactions; however at high temperature, both the GSKS $v_{xc}$ approximation becomes less suitable and the thermal energy dominates over many-body interactions,  leading to more accurate results for the hybrid NI approximation.

\section{Discussion}
\subsection{`Hybrid' versus `simple' approximations}
\label{simpvshybr}

From the results presented here  and in \cite{Skelt2019JPA} it is clear that all hybrid approximations improve substantially over the correspondingly simple approximations.
To understand the reason for this, we turn once more to the spectra in figure~\ref{fig:simple_spectrum}.
As $U$ increases, at small $t$, the main features of the exact instantaneous spectrum are going to be determined by many-body interactions; however, at large $t$ and up to relatively large $U$'s they remain mainly dominated by the driving potential, which is single-particle. These non-interacting features, are relatively well reproduced by all the approximations, and in particular, with the exception of BALDA at large $U$ values, the energy range spanned by the approximate spectra at $t\approx\tau$ remains relatively close to the exact one. At $t/\tau\approx 0$, approximations fail to reproduce the charge-gap which opens for increasing $U$ (NI and GSKS) and/or quantitatively fails to reproduce the energy spread of the exact instantaneous spectrum, tending to underestimate it. This reflects in a density of states at $t/\tau\approx 0$ which is increasingly different from the exact one for increasing $U$ (compare black lines in  figure~\ref{fig:DoS}) 
As a result, in the simple approximations, at $t=0$ the thermal
energy tends to populate the wrong number of states, too few or
too many, does not matter.  
As we are considering closed systems, no decay may help adjusting the DOS during the dynamics  and the wrong number of states will tend to be
populated at $t\approx\tau$, where the spectra (and the DOS) tend to be more similar to the exact one (figure~\ref{fig:simple_spectrum} and purple lines in  figure~\ref{fig:DoS}), respectively. 
Hence the extracted work will be substantially different from its exact counterpart, with the initial state occupation strongly affecting the number and weight of the significant contributions to $P(W)$.

The hybrid approximations provide correction to the initial state occupation. In this way, large part of the negative effect of the instantaneous spectrum non-reproducing (GKSK) or reproducing only qualitatively (BALDA) the charge-gap of the Mott-insulator transition is overcome. In particular, because, for the types of potentials explored, at large $t$ the approximate and exact instantaneous spectra are similar, by correcting occupation at small $t$'s, the hybrid approximations strongly improve results in the quasi-sudden quench regime.

We note that in any of the approximations presented the effective number of energy levels is reduced in comparison to the spectrum of the exact Hamiltonian as many degeneracies can be lifted only by an Hamiltonian that is explicitly interacting. This can be noted by comparing the spectra and DoS's in Figs.~\ref{fig:simple_spectrum} and \ref{fig:DoS} respectively.  These degeneracies reflect in the approximate $P(W)$'s which will comprise fewer distinct transitions.

\subsection{Overall comparison of results for the approximated average work}

\begin{figure*}
\centering

\includegraphics[width=1\textwidth]{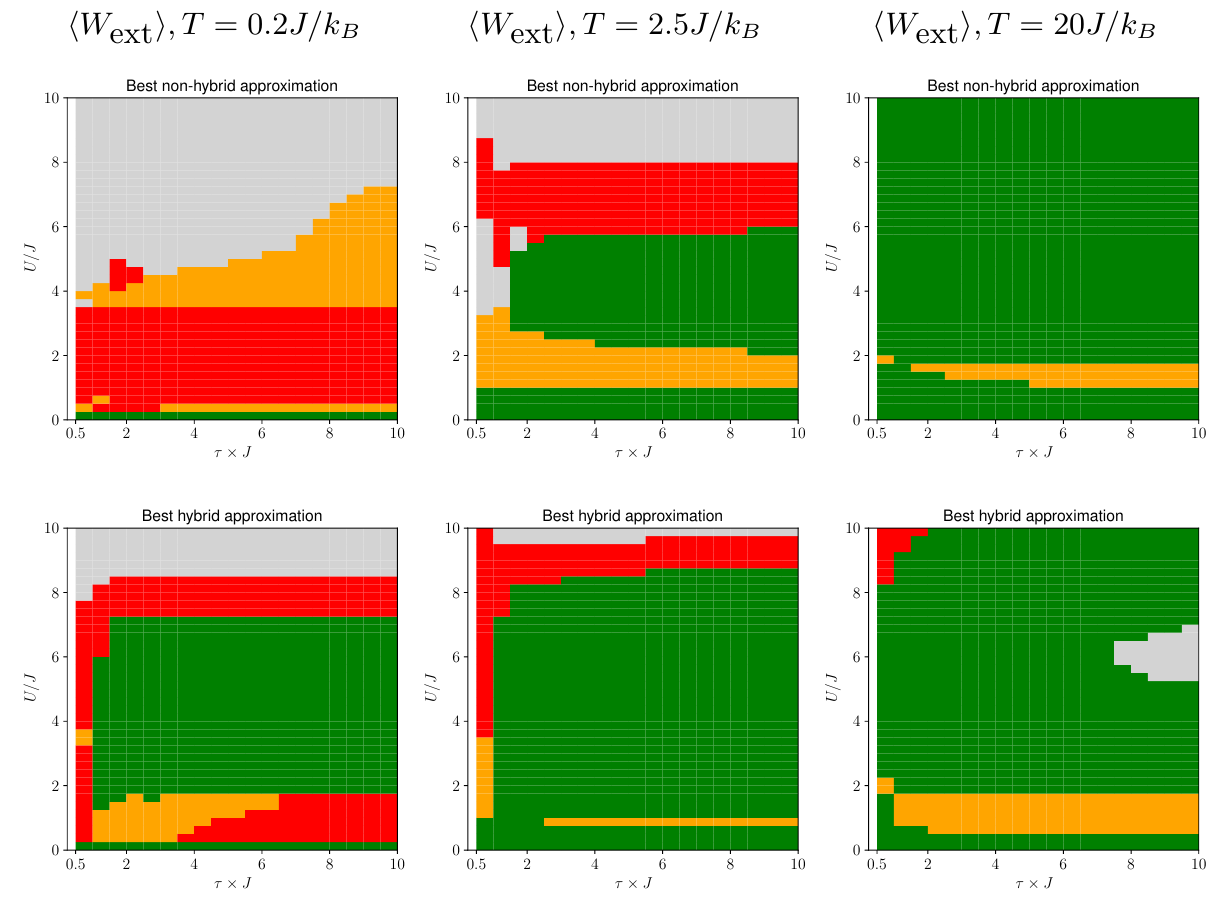}
\caption{Upper panels: Figures showing which simple approximation is most accurate (up to 20\% relative error) for the average work in each region (6 site chains, `comb' potential).The colours representing the approximations are green (NI), orange (BALDA), and crimson (GSKS).  Grey is used when all approximations have a relative error above 20\% . Lower panels: as for the upper panels but for the hybrid approximations. All approximations reduce to the simple NI for $U=0$.
}
\label{fig:best_approxes_work}
\end{figure*}

We summarize comparisons among all approximations in Figure~\ref{fig:best_approxes_work} where we consider accuracy down to at most $20\%$, and report which approximation performs best within this constraint at every point in the parameter space. The top row shows the outcome for the average quantum work for all `simple' approximations at low, medium, and high temperature (left to right); the bottom row shows the corresponding results for the `hybrid' approximations.
The colour code in the figure is: 
\begin{itemize}
    \item \colorbox{limegreen}{Green} when the simple/hybrid NI approximation has the best accuracy; 
     \item \colorbox{orange}{Orange} when the simple/hybrid BALDA has the best accuracy;
    \item \colorbox{crimson}{\textcolor{white}{Crimson}} when the simple/hybrid GSKS approximation has the best accuracy; 
    \item \colorbox{lightgray}{Gray} when the relative error for all approximation is over $20\%$;
    \item \colorbox{purple}{\textcolor{white}{Purple}} when all approximations are at least $20$\% accurate; 
\end{itemize}

At low temperatures and for the `simple' approximations, including interactions via  an accurate, system-tailored, low-temperature (albeit static) $v_{xc}$ (in addition to a static Harthree potential) gives consistently the best approximation up to medium many-body interactions $U\stackrel{<}{\sim} 4$. This is indeed the regime in which we would expect the GSKS approximation to best behave. Interestingly `BALDA' outperforms it  at higher $U$ values when in the quasi-adiabatic regime (high $\tau$). Here the system would be slowly varying in time and higher values of $U$ confine the dynamics mostly to the evolution of the instantaneous ground state. From figure~\ref{fig:simple_spectrum} it can be then see that the energy difference between initial and final ground states is approximated much better by BALDA than by GSKS. In this regime, this is the major contribution to the work. The simple BALDA approximation here used can be considered as a zero-order implementation of the adiabatic-LDA\cite{Ullrich2013}
which is indeed designed for zero-temperature dynamical systems in the adiabatic regime.

At intermediate temperatures, simple GSKS approximation is the best in the regime for which Coulomb repulsion is dominant over the thermal energy and starts to reduce access to excited states with double occupation. While GSKS instantaneous spectrum does not reproduce the charge-gap due to the Mott-insulator transition, its low-energy level spacing allows for occupation of the relevant eigenstates (improvement over BALDA) and contain some information about Coulomb repulsion (improvement over NI). At lower $U$ values Coulomb and thermal energy become comparable and for most of the parameter regions  the simple NI approximation returns the best results, and similarly for the great part of the parameter region at high temperatures.  There are regions at low $U$ values where the simple BALDA is better than the simple NI: a medium-high temperature will reduce both the variation in the static density and the response to the external applied field providing a density profile more appropriate to a BALDA-type approximation. However we think that the boundary of the region in which simple BALDA behaves the best are somewhat accidental, especially at high temperature, where the exact average work varies little with the parameters and hence any approximation giving a result in the correct ballpark has a chance to be `the best'. At these high-temperatures, the dominance of the thermal energy over Coulomb interaction and external field suggests the simple NI approximation to be the most trustworthy.

Inclusion of many-body interaction effects through the initial exact thermal state remains the single most important improvement with respect to the simple NI approximation. In fact, Figure~\ref{fig:best_approxes_work} confirms that the hybrid NI approximation is the best performing approximation in the majority of parameter regimes at all temperatures.  The hybrid GSKS outperforms the hybrid NI primarily in the strongly correlated  and quasi-sudden-quench regimes. In the first case many-body interaction effects in $H^{\evo}$ help maintaining precision where Coulomb repulsion dominates; in the second case, the improvement over the instantaneous initial and final spectra delivers a better approximation. The hybrid BALDA approximation occasionally performs as best in weakly-coupled regimes at the low and high temperatures. Here, though, also the hybrid GSKS performs within $10\%$ of the exact results.

All approximations reduce to the simple NI for $U=0$.

\begin{figure*}
\centering

\includegraphics[width=\textwidth]{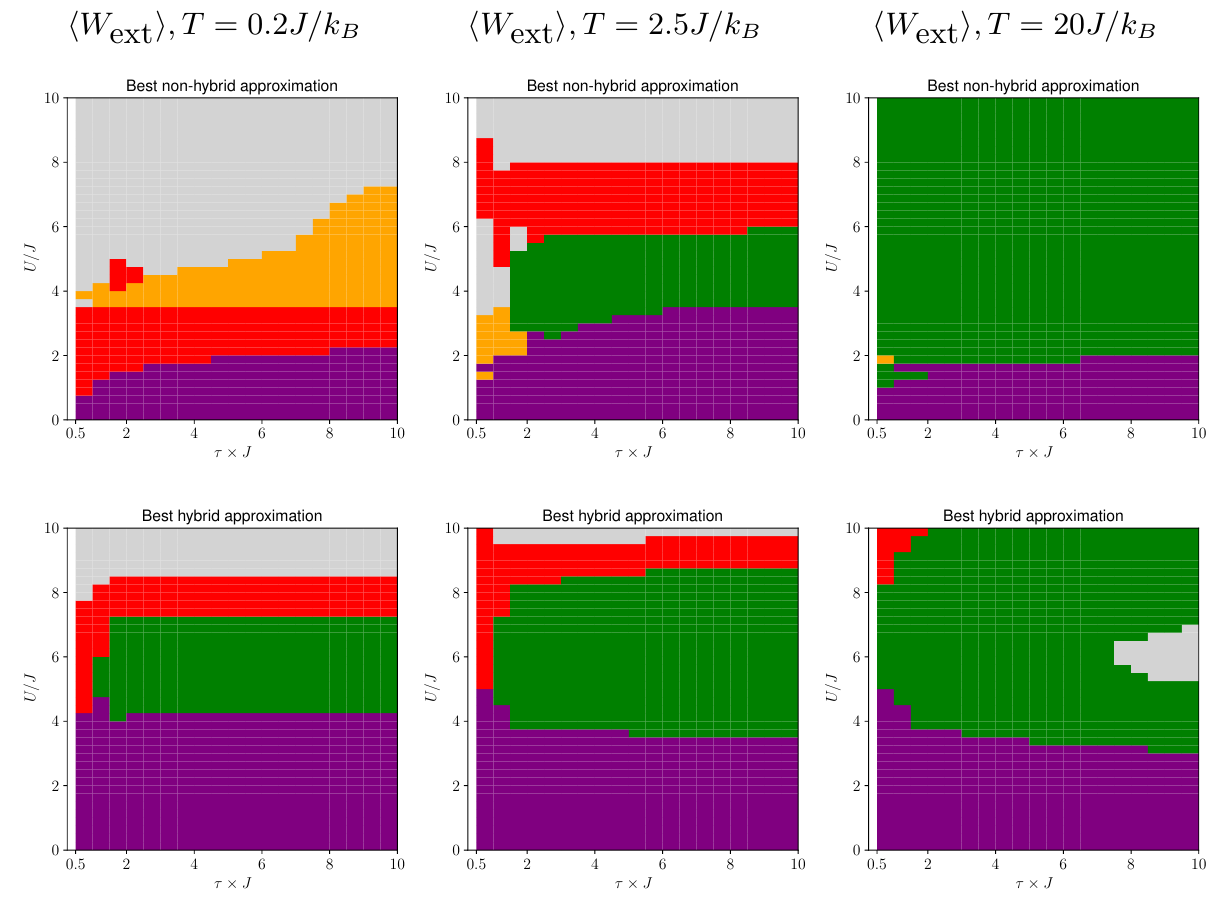}
\caption{As for figure~\ref{fig:best_approxes_work}, but with purple color representing where all approximations are accurate within 20\%
 relative error.}
\label{fig:best_approxes_work_all_best}
\end{figure*}

In some circumstances, one may prefer to choose a single approximation, provided that its error is below a given threshold for the parameter region of interest. To aid with this, in figure~\ref{fig:best_approxes_work_all_best} we re-plot the results of Figs. \ref{fig:best_approxes_work} highlighting in purple the regions of the parameter space in which all approximations yield less than 20\% of error.

\subsection{Overall results for the approximated entropy variation}
\label{sec:results_entropy}

We construct approximations to the entropy variation equation~(\ref{eq:entropy_ext}) similarly to reference~\cite{Skelt2019JPA}: the approximation used to calculate the free energy variation matches the approximation used for the initial state of the system, and $\Delta S_{approx}=0$ if $\left( \langle W_{ext}^{appr} \rangle + \Delta F^{appr} \right)>0$ to avoid nonphysical negative entropy variation. In this case  $|\Delta S_{exact}-\Delta S_{approx}|/|\Delta S_{exact}|=1$ confirming that in that case the approximation is not a good one.

From the exact results in Fig. 2 (central column) and Fig. 3 (central panel, first row) of  ~\cite{Skelt2019JPA}, we observe that $\langle W_{ext}^{exact} \rangle$ is always positive and $\Delta F^{exact}$ is always negative for the `comb' potential. In addition, they are often very close in absolute value. This implies that $\Delta S$ is often the result of a cancellation of terms. Then, to obtain a good approximation for the entropy variation, it is necessary that not only work and free energy are both separately well approximated, but also that their inaccuracies combine point-by-point in the parameter space so to keep the overall error low. This is not a given.

Indeed, the simple NI and hybrid NI approximations reproduced the entropy quite poorly and inconsistently for most regimes and temperatures, see figures B1 and B2 of  ~\cite{Skelt2019JPA}. Here we wish to explore if including many-body interactions through a KS potential in the evolution Hamiltonian improves the reliability of these types of approximations, at least for certain interaction regimes.

\begin{figure*}
\centering

\includegraphics[width=\textwidth]{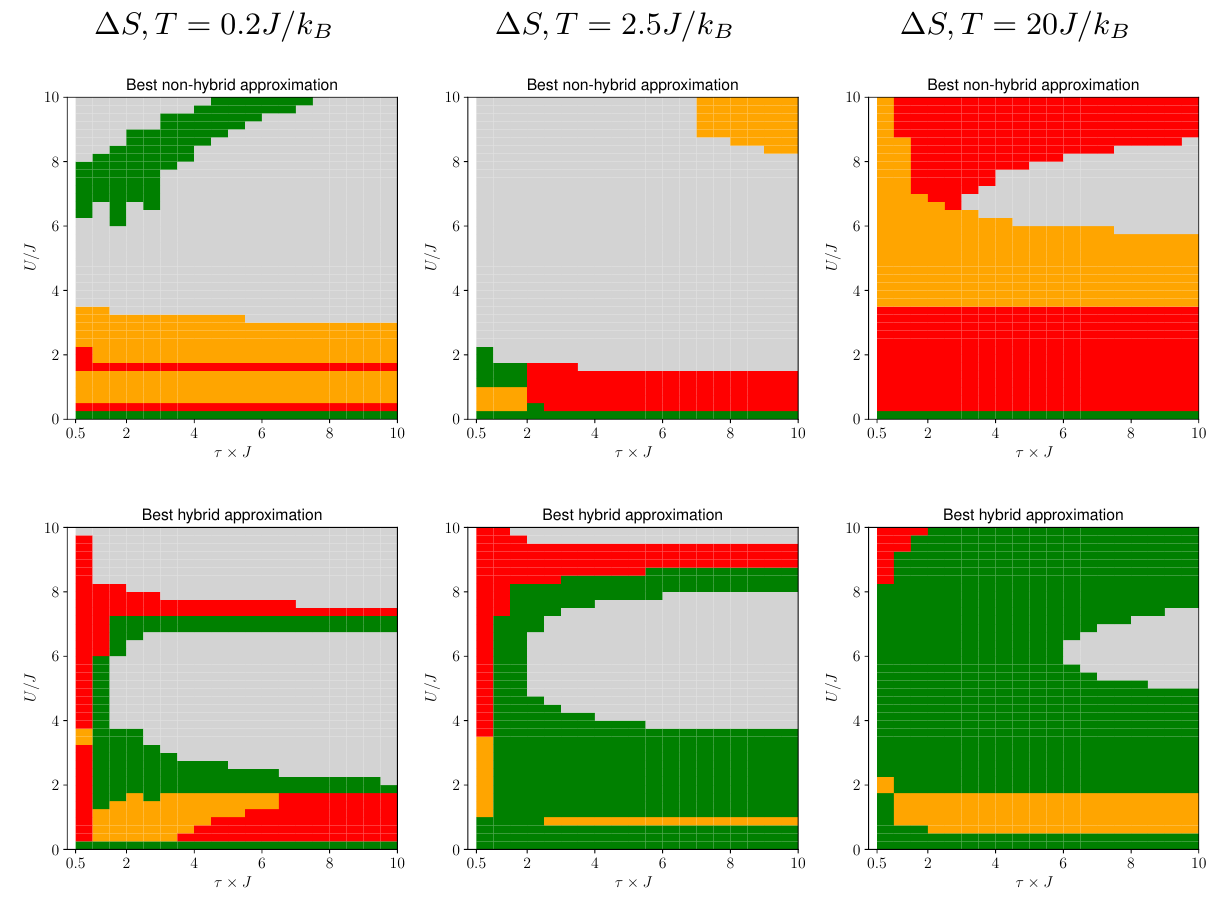}
\caption{Same as in figure~\ref{fig:best_approxes_work} but for the entropy variation.
 }
\label{fig:best_approxes_entropy}
\end{figure*}

We start by discussing the `hybrid' approximations (lower row of figure~\ref{fig:best_approxes_entropy}). When $\left( \langle W_{ext}^{appr} \rangle + \Delta F^{appr} \right)<0$, the relative error on the entropy variation is related to the corresponding error on the extracted work as follows
\begin{equation}
    \frac{|\Delta S_{exact}-\Delta S_{approx}|}{|\Delta S_{exact}|} =
    R_{exact} \frac{|\braket{W^{exact}_{ext}}-\braket{W^{approx}_{ext}}|}{|\braket{W^{exact}_{ext}}|},
\end{equation}
 with
\begin{equation}
  R_{exact} = \frac{1}{k_B T}\frac{|\braket{W^{exact}_{ext}}|}{|\Delta S_{exact}|}.
\end{equation}
This shows that only where $R_{exact}\le 1$ the hybrid approximation for the entropy variation is at least as good as the corresponding one for the average work.   Importantly, as in this case $\Delta F$ is exact, an improvement over the approximation of $\langle W_{ext} \rangle$ should directly reflect on the quality of the approximation for $\Delta S$.
Indeed the results obtained from the hybrid GSKS approximation complements results from hybrid NI approximation in a similar way as for the results for the work, providing accuracy especially in the sudden quench and strong interaction regimes. 

The comparison of the `simple' approximations is presented in the upper row of figure~\ref{fig:best_approxes_entropy}. Here the DFT-based approximations consistently provide good estimates for $\Delta S$ in the low-interaction regime. Results are particularly good at high temperatures. In this regime the $\Delta S$ is small and so is its variation over the parameter range (less than a factor 2). Both simple GSKS and simple BALDA reproduce the highest value of $\Delta S$; simple BALDA strongly underestimates $\Delta S$ lowest value, though reproducing the qualitative behaviour of the exact entropy variation better than GSKS. However both approximations are designed for low temperatures, so further study would be necessary to understand if the accuracy displayed is just a coincidence.

The detailed results for the simple and hybrid GSKS and BALDA approximations are reported in \ref{appendix}. Information on parameter regions in which all approximations have a relative error of at most 20\% is in figure~\ref{fig:best_approxes_entropy_all_best}.  

\section{Conclusions}

Calculating non-equilibrium properties of driven finite-temperature many-body systems is at the core of quantum thermodynamics but remains a hard task.
Here we introduced DFT-inspired approximations to the calculation of the average work and entropy variation. 
These approximations can be applied to generically-driven, large many-body systems. We exemplify their use by considering one-dimensional Fermi Hubbard chains, driven to display transitions between different phases: metal, Mott-insulator, and band-insulator phases. 

Building on 
the structure of the approximations in references~\cite{Herrera2017,Herrera2018,Skelt2019JPA}, we (further) included many-body effects by approximating the driving many-body Hamiltonian with Kohn-Sham Hamiltonians. These included  static Hartree and exchange-correlation potentials. This improves sophistication while remaining computationally cheap. 
We introduced the `hybrid' versions of the local density approximation (BALDA) and of the exact (reverse-engineered) ground state Kohn-Sham potential  (GSKS) for the average work and entropy production.  We compared these hybrid approximations to their equivalent `simple' approximations, as well as comparing the results to the corresponding non-interacting approximations from reference~\cite{Skelt2019JPA}.%

`Hybrid' approximations require the ability of exactly (or very accurately) diagonalize the initial (for the work) and also the final (for the entropy variation) Hamiltonians, but are confirmed to strongly improve results of corresponding `simple' approximations.

`Simple' BALDA and GSKS are designed for low-temperature systems. Indeed, when considering the average work, they are the best `simple' approximations in that regime and even for some parameter regions at intermediate temperatures. Further, all the `simple' approximations considered have better than 20\% relative error in the weakly interacting regime at all temperatures.  

When considering `hybrid' approximations for the average work, all approximations behaves well (below 20\% error) at all temperatures for low and medium interaction strength. The non-interacting  `hybrid' approximation remains the best though in most parameter regions, with the exception of sudden quench and highly interacting regimes, where the use of an accurate initial KS potential pays off. In general, `hybrid' approximations should provide significant improvement when the system evolves towards a similarly or lesser-interacting regime, as they allow for many-body corrections to the initial thermal population of the system's state. The initial population may be otherwise quite flawed as the gaps' structure of the spectra of the Kohn-Sham Hamiltonians is very different from the many-body one.

Performance for the entropy variation follows similar trends to the average work, though areas with relative error greater than 20\% are larger at all temperatures. The entropy variation is generally small and the result of a cancellation of terms: we believe that this contributed to the larger relative error.  

To further improve the approximations' performance, we would  recommend to include dynamic corrections e.g. by using TDDFT-style approximations for the driving Hamiltonian, such as in reference~\cite{Herrera2018}.  However this is a much more computationally expensive scheme. 

An LDA approximation or a reverse engineering scheme dealing specifically with systems at finite temperature could also be considered. For instance an adaptation of BALDA has been proposed in reference~ \cite{Vivaldo-PRA.92.013614}. Employing the formalism of thermal DFT will be the focus of further studies.

\subsection{Acknowledgments}
 AHS thanks EPSRC for financial support. KZ thanks the S\~ao Paulo Funding Agency FAPESP grants 2016/01343-7 and 2020/13115-4.

\section*{References}
\bibliographystyle{iopart-num}
\bibliography{newref}

\appendix 
\section{Results for the approximated entropy variation} \label{appendix}

Below, we show results for the entropy variation $\Delta S$ from equation~\ref{eq:entropy_ext} and the corresponding relative error with respect to the exact system. The parameter space as well as the three temperatures considered are the same as for the average work.
The darker the orange shade, the less entropy is produced, and the deeper the purple shade, the more accurate the quantitative results are.

\label{Apx:performance_all_approximations}
\subsection{`Simple' BALDA}

\begin{figure*}
\centering

\includegraphics[width=0.95\textwidth]{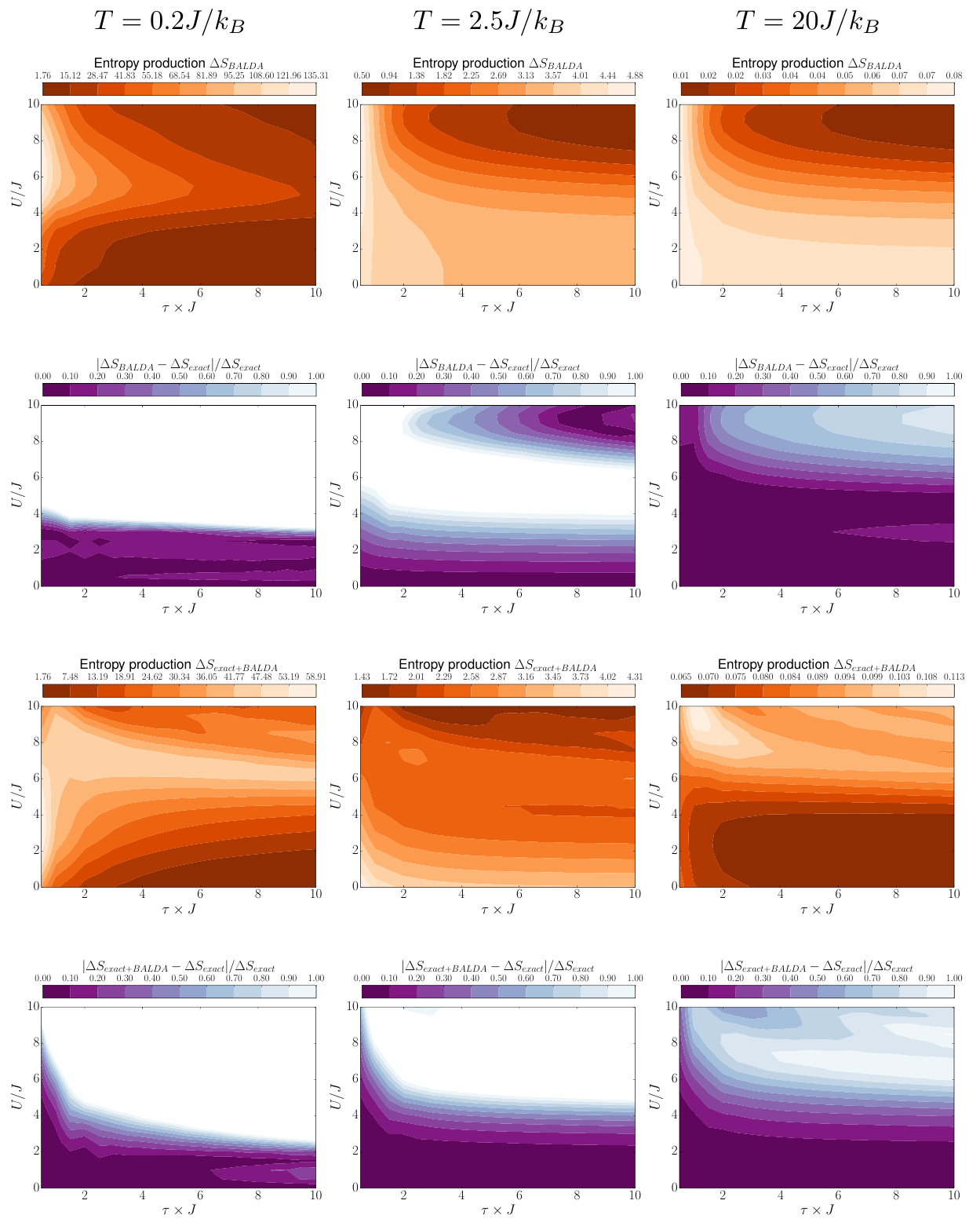}

\caption{First row: Entropy produced using `simple' BALDA for $0.5 \leq \tau \times J \leq 10$ ($x$-axis) and $0 \leq U/J \leq 10$ ($y$-axis) for 6 site chains with `comb' potential, increasing temperature from left to right.
Second row: Relative difference between the `simple' BALDA entropy production and the exact results for the same parameters as the upper panels. Third and fourth rows: same as in first and second row, but for `hybrid' BALDA.
}
\label{fig:BALDA_entropy}
\label{fig:ex+BALDA_entropy}
\end{figure*}

Figure~\ref{fig:BALDA_entropy} shows results for the entropy within the `simple'-BALDA approximation. 

Overall, the `simple' BALDA entropy production is larger for sudden quenches, while decreasing with temperature. Qualitatively it reproduces well the exact behaviour at low and high temperature, but less so at intermediate one. Quantitatively, it  performs better at high temperatures, with less than 10\% of error within a large portion of the parameter space. 

\subsection{`Hybrid' BALDA}

Results for the `hybrid' BALDA approximation are shown in  figure~\ref{fig:ex+BALDA_entropy}.

Interestingly, this hybrid approximation fails to reproduce qualitatively the exact behaviour at all temperatures. We speculate that this is due to the two terms on the r.h.s. of equation~\ref{eq:entropy_ext} been approximated in different ways, so that even if each approximation independently has the correct qualitative behaviour, this is lost in their overall sum. 

The overall range of the entropy production is well captured at all temperatures.

\subsection{`Simple' GSKS}

\begin{figure*}
\centering
\includegraphics[width=0.95\textwidth]{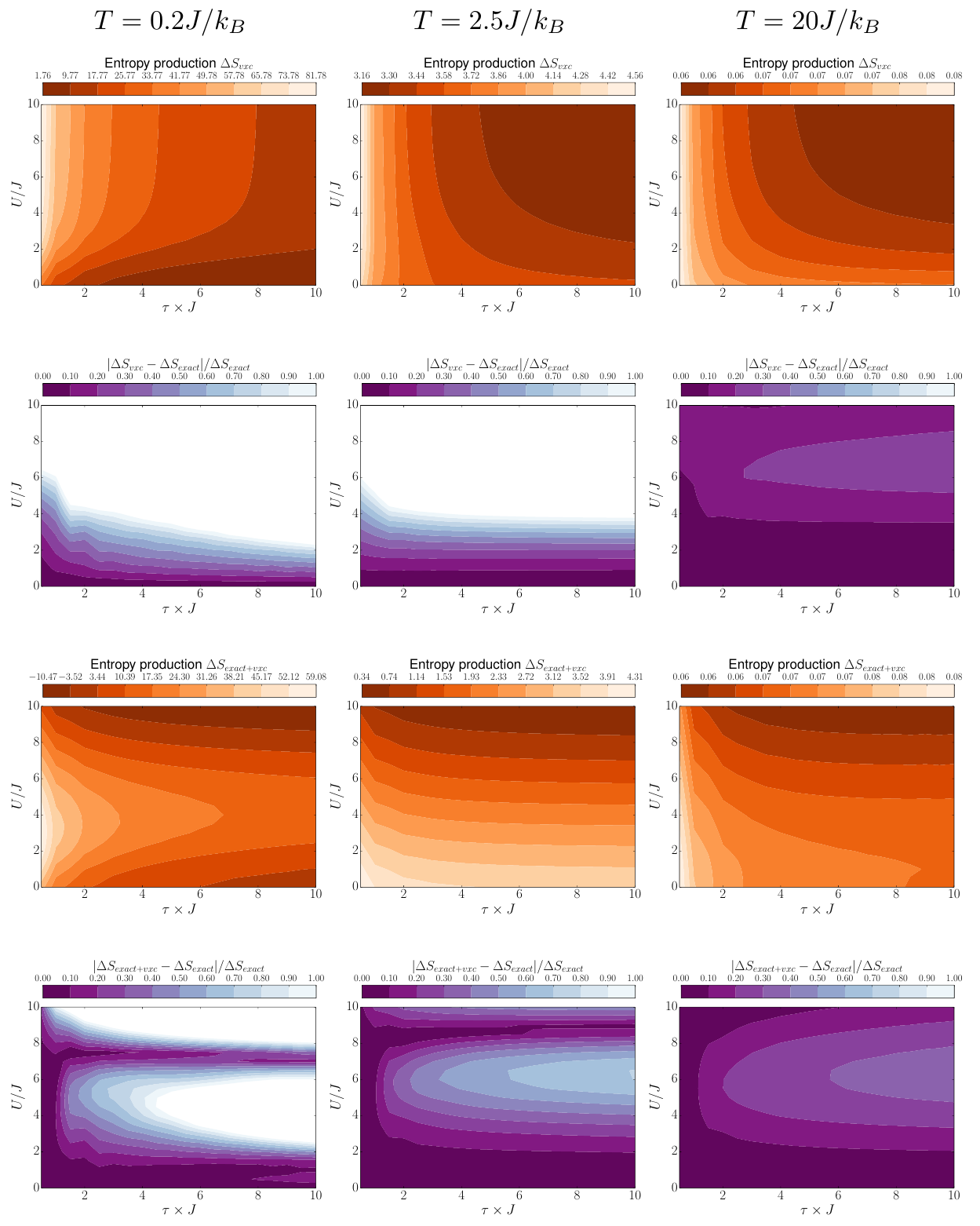}
\caption{Same results and parameters as in figure~\ref{fig:BALDA_entropy} for GSKS `simple' and `hybrid' approximations.}
\label{fig:vxc_entropy}
\label{fig:ex+vxc_entropy}
\end{figure*}

Figure~\ref{fig:vxc_entropy} shows results for the `simple' GSKS approximation.

Entropy production at high temperatures is well approximated, this time with substantial improvement over simple and hybrid BALDA, but not at intermediate and low $T$. The highest error is of the order of 30\%. At low and medium $T$,  however, we observe a poor performance mainly due to errors in the free energy.
Overall the qualitative behaviour does not reproduce well the exact one.

\subsection{`Hybrid' GSKS}

Figure~\ref{fig:ex+vxc_entropy} depicts results for the entropy variation calculated using the `hybrid' GSKS approximation. 

Qualitatively, an overall behaviour similar to the exact one is recovered at low and intermediate temperatures.

Quantitatively, accuracy is achieved in almost all parameter space at the high temperature; we also see an improvement in the results for low and intermediate temperatures over all other approximations considered (simple and hybrid NI/BALDA and also simple GSKS).

\subsection{Overall comparison}

	In some circumstances, one can be interested in choosing a single approximation that yields a given error threshold.
	If we modify the criterion discussed in Sec. \ref{subsec:comparison} to account for regions of the parameter space in which all approximations yield less than 20\%, we obtain the following results.

\begin{figure*}
\centering\includegraphics[width=\textwidth]{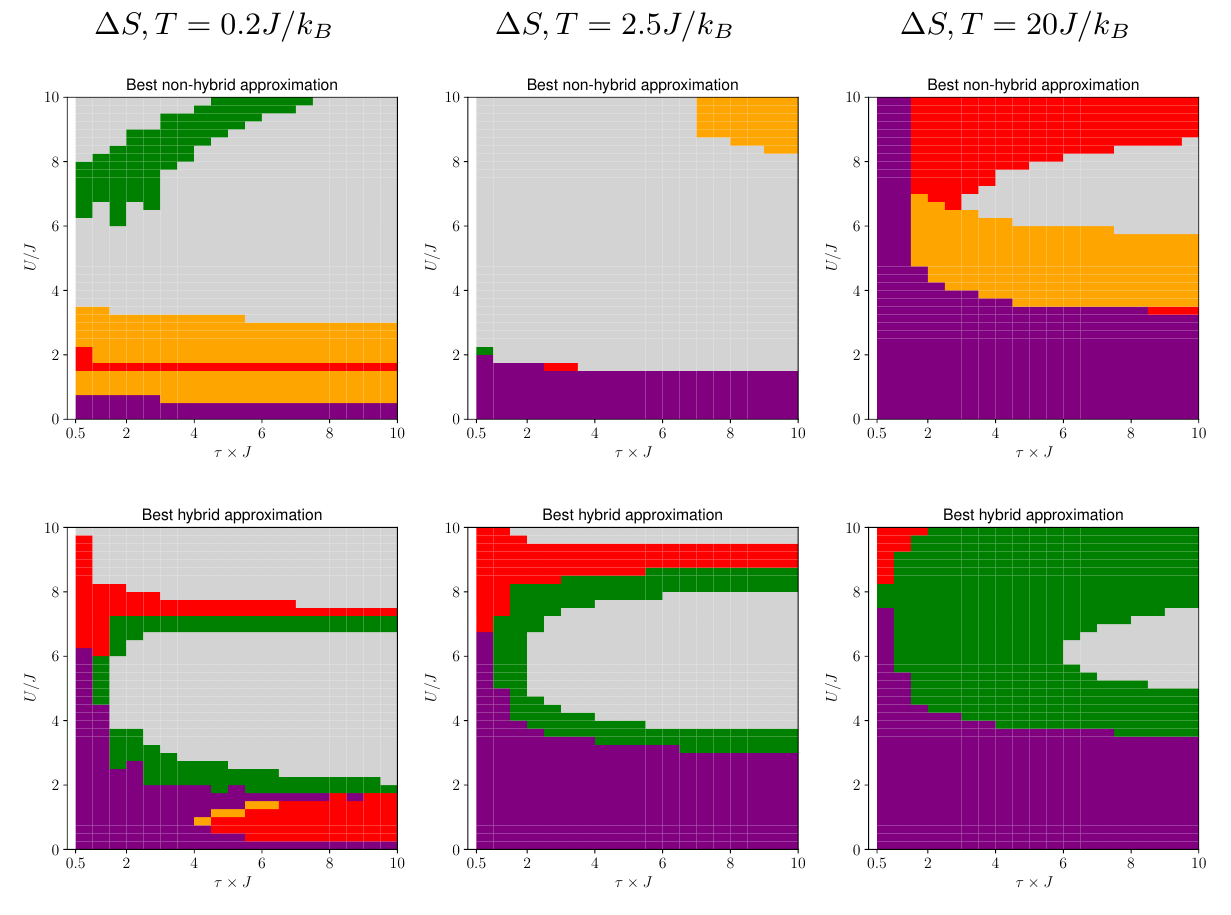}
\caption{ Upper panels: Figures showing which `simple' approximation is most accurate (up to 20\%) for the entropy production. 
The colours representing the approximations are green (NI), orange (BALDA), and crimson (GSKS), and purple (all).  Grey is used when all approximations have a relative error above 20\% . Lower panels: as for the upper panels but for the hybrid approximations. All approximations reduce to the simple NI for $U=0$.
Lower panels: Figures showing which hybrid approximation is most accurate (up to 20\%) for the entropy production. 
}
\label{fig:best_approxes_entropy_all_best}
\end{figure*}

\section{Effect of thermal density on BALDA-type approximations}
\label{app_BALDA_temp}

\begin{figure*}
\centering
\includegraphics[width=0.9\textwidth]{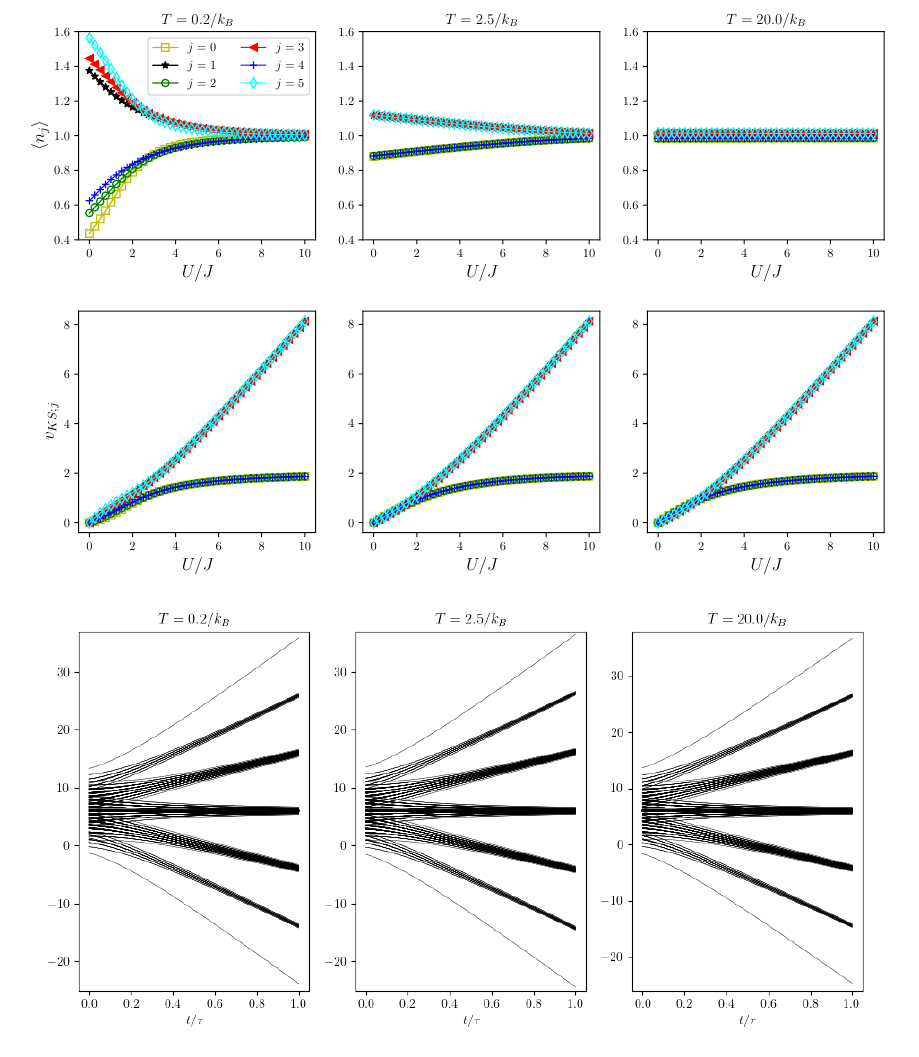}
\caption{ First row: Exact $t=0$ local thermal densities for $T = 0.2J/k_B$, $T = 2.5J/k_B$ and $T = 20J/k_B$ (left to right). The densities are plotted as a function of $U/J$. Different colors correspond to different sites. 
Second row: BALDA $t=0$ KS potentials corresponding to the densities in the first row . The potentials are plotted as a function of $U/J$. 
Third row: $U/J=2$ simple and hybrid BALDA spectra stemming from the evolution of the potentials in the second row as a function of $t/\tau$. 
}
\label{BALDA_ini_vKS}
\label{BALDA_ini_n}
\label{BALDA_spectra} 
\end{figure*}

In figure~\ref{BALDA_ini_n}, first row, we show the $t=0$ thermal densities used for constructing the KS potentials $v^{BALDA}_{KS,i}$ for the BALDA approximation for all temperatures considered as a function of $U/J$, Each color labels a site.

In figure~\ref{BALDA_ini_vKS}, second row, we show the $t=0$ KS potential for the simple and hybrid BALDA approximations, for all temperatures considered, and as a function of $U/J$. Each color labels a site.

In figure~\ref{BALDA_spectra}, third row, we show the $U/J=2$ instantaneous spectra for the simple and hybrid BALDA approximations, for all temperatures considered, and as a function of $t/\tau$.  

\end{document}